# $10^{-15}$-level laser stabilization down to fiber thermal noise limit using self-homodyne detection

IGJU JEON, WOOSONG JEONG, CHANGMIN AHN, AND JUNGWON KIM*

*Korea Advanced Institute of Science and Technology (KAIST), Daejeon 34141, Republic of Korea*
**jungwon.kim@kaist.ac.kr*

**We demonstrate a self-homodyne detection method to stabilize a continuous-wave 1550-nm laser to a 1-km optical fiber delay line, achieving a frequency instability of $6.3\times10^{-15}$ at a 16-ms averaging time. This result, limited by fiber thermal noise, is achieved without the need for a vacuum system, highlighting the potential of our approach for ultra-stable laser systems in non-laboratory environments. The system utilizes only a few passive fiber optic components and a single balanced photodetector, significantly simplifying the laser stabilization process while maintaining high performance. The entire optical setup is compactly packaged in a portable metal air-tight case.**

There is an increasing demand for ultra-stable laser sources that can be used in non-laboratory environments such as atmospheric and gas sensing [1,2], transportable lattice clock [3,4], and earthquake (moonquake) detection [5–7]. Ultra-stable optical frequency comb sources operating outside laboratories can be also utilized in field-deployed dual-comb spectroscopy [8,9] and generation of low-noise microwave [10,11]. To achieve an ultra-low-noise laser system, the free-running laser needs to be stabilized to a mechanical length reference using a suitable frequency discrimination system, where both the reference and discriminator should be compact and robust against external mechanical disturbances.

Recent compact laser stabilization approaches focus on small-sized optical cavities with ultra-high Q factors and low thermal noise. Examples include ultra-low expansion (ULE) Fabry-Pérot cavities, dielectric whispering-gallery-mode (WGM) resonators, integrated chip-scale resonators, and optical fiber delay-lines. For instance, vacuum-gap Fabry-Pérot [12–14] and air-gap cavities [15,16] have achieved frequency instabilities of $10^{-14}$ to $10^{-15}$, though they require free-space alignment. Dielectric-WGM resonators [17] or microrod-based systems [18] have achieved stability levels ranging from $10^{-14}$ to $10^{-13}$, which are limited by the thermal noise of the reference. Integrated chip-scale resonators, despite achieving $10^{-13}$ stabilization, face limitations due to higher thermal noise limit and frequency drift [19].

These systems often rely on the Pound-Drever-Hall (PDH) method [20], which involves optical modulators and many RF components. An alternative is fiber-based systems, such as stimulated Brillouin scattering (SBS), which can achieve $10^{-13}$ instability using only 2-m fiber coil [21,22]. Although the SBS process itself does not require the PDH system due to its passive nature, practically, the PDH method is needed to tune the pump laser frequency to the cavity resonance.

Longer optical fiber delays, using self-heterodyne interferometers [23–29] or ring-resonator configurations [30], offer a compact, alignment-free solution. These can stabilize both CW lasers [24,28–30] and optical frequency combs [25–27], achieving high performance with simpler setups. The self-heterodyne systems using kilometers of fiber delay have demonstrated remarkable stability, with sub-Hz linewidths and frequency instabilities as low as $3.2\times10^{-15}$ at 1 second [29]. However, achieving these results relies on thermal shielding and vacuum systems specifically designed to enhance long-term stability, which introduce additional design considerations and increase the overall system size. Furthermore, advances in fiber optic coil winding have enabled compact fiber spools, yet in self-heterodyne configurations, components like acousto-optic frequency shifter still limit system miniaturization [27]. Alternatively, a recent solution involves an all-passive ring-resonator with 100-meter polarization-maintaining (PM) fiber, which has achieved a Q factor of 4 billion and $10^{-14}$-level frequency instability within 0.1 seconds [30]. This design effectively suppresses intensity noise but requires all-polarization-maintaining fiber components and careful management of cavity loss to maintain a high Q factor.

In this Letter, we adopt an all-single-mode (SM) fiber self-homodyne interferometric configuration with a 1-km vibration-insensitive optical fiber delay to stabilize CW lasers. Unlike self-heterodyne interferometers, which require acousto-optic modulators and additional RF components, this system uses only a few passive optical components and a single balanced photodetector (BPD) to generate the error signal. The 1-km fiber delay with the Michelson interferometer offers a Q factor of 4 billion. A simple BPD can effectively suppress the coupling of intensity noise of the laser to be stabilized to the frequency noise. We stabilized CW lasers to a $10^{-15}$-level frequency instability, where the frequency noise is limited by fiber thermal noise in the 4 Hz to 200 Hz range.

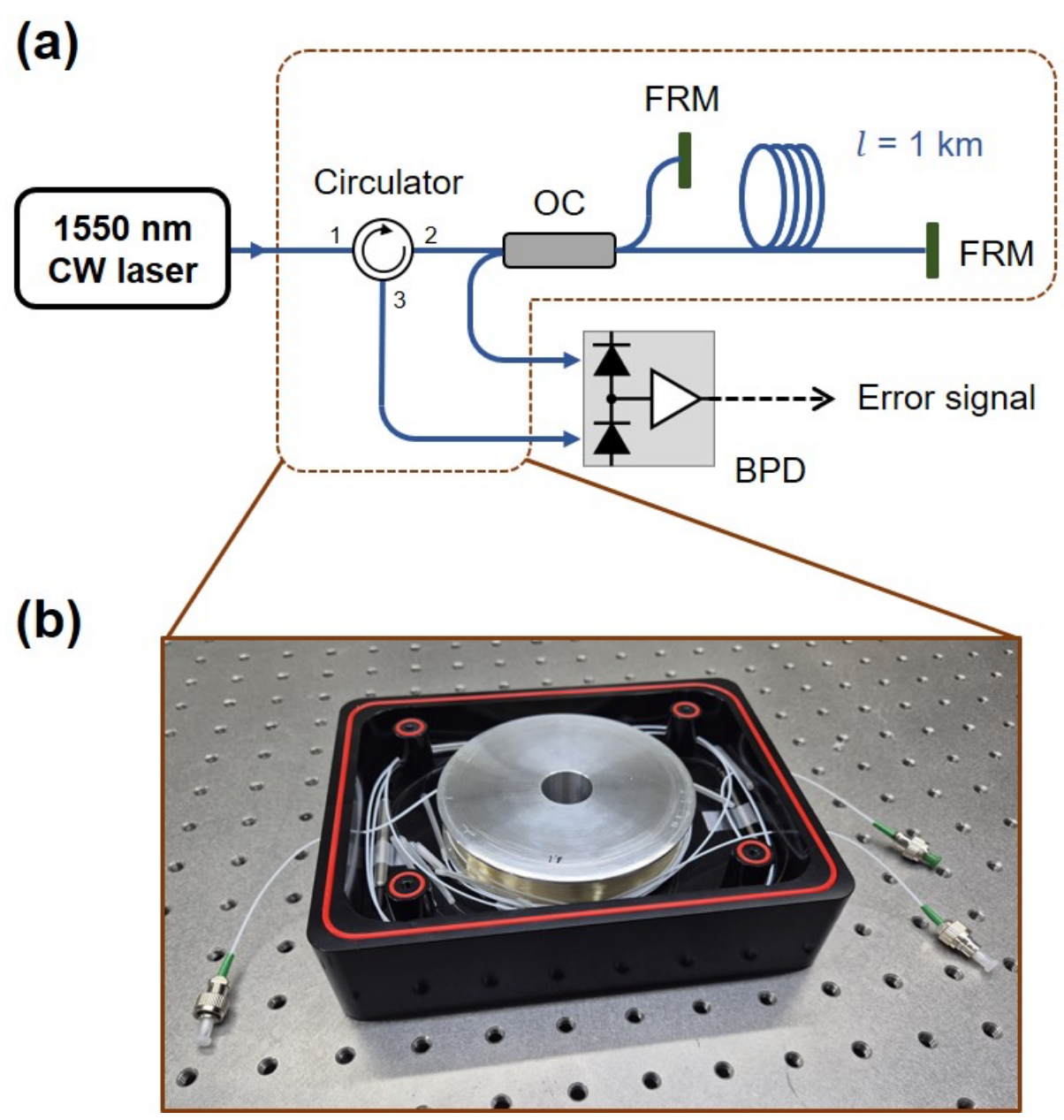

**Fig. 1** (a) configuration of the all-fiber self-homodyne Michelson interferometer with balanced photodetector. OC, optical coupler; FRM, Faraday rotating mirror; BPD, Balanced photodetector. (b) Packaging of the self-homodyne Michelson interferometer with vibration-insensitive 1-km optical fiber spool. The dimension of the case (with lid) is 140×180×51 mm$^3$. The photograph does not include the 10-mm thick lid.

The fiber-delay-based Michelson interferometer is composed of four main components: a fiber-coupled circulator, a balanced optical fiber coupler, an optical fiber delay, and two fiber-coupled Faraday rotating mirrors, as shown in Fig. 1(a). The laser to be stabilized enters port 1 of the circulator and exits from port 2, proceeding to the fiber interferometer. We used two Faraday rotating mirrors at the end of the interferometer arms instead of standard fiber-coupled mirrors to match the polarization state between the two interfering signals, maximizing the signal-to-noise ratio (SNR) of the interference signal. To suppress voltage signal fluctuations caused by laser intensity noise, the two interfering optical signals are subtracted using balanced photodetection. One signal comes directly from the coupler, while the other enters port 2 of the circulator and exits from port 3.

Unlike other platforms where resonators with ultra-high Q factors have narrow dips separated by the free spectral range (FSR), this configuration consistently features a spectral width equal to half of the FSR, with a simple sine-wave-shaped transmittance along the frequency of the laser. Since the FSR is $c/nl$ where $c$ is the speed of light, $n$ is the refractive index of the fiber, and $l$ is the effective length of the fiber delay, the spectral width of the configuration is $c/2nl$. Note that since we used a Michelson interferometer instead of a Mach-Zehnder interferometer, the effective length of the reference is twice the actual length of the fiber spool we used.

The Q factor of the configuration can be computed as $4\times10^9$ by dividing the optical carrier frequency (193.5 THz at 1550 nm wavelength) by the spectral width. This configuration allows us to achieve an ultra-high Q factor, which can effectively discriminate laser frequency noise. Compared to previous research on all-fiber ring-resonator configurations [30], this setup does not require PM fiber components. Additionally, the Q factor of the system is maintained regardless of losses inside the cavity.

We packaged the self-homodyne Michelson interferometer, based on a vibration-insensitive 1-km optical fiber spool, in a compact metal enclosure. As shown in Fig. 1(b), the setup uses off-the-shelf optical components arranged as in Fig. 1(a) and includes an optical fiber spool within an air-tight aluminum case measuring 140×180×51 mm$^3$. The case has an input for the laser signal that needs stabilization and two optical outputs for balancing. To achieve air-tight packaging, we used rubber seals and screws, effectively preventing acoustic noise from compromising the stability of the fiber reference. The packaged setup was used in the stabilization experiments described later.

Figure 2 shows the schematic of the laser stabilization and performance measurement setup. Two 1550 nm diode lasers (ULN15PT and ULN15TK, Thorlabs Inc.) are used as CW laser sources. Approximately 1 mW of optical power from each laser is directed into independent frequency discriminators, each configured as shown in Fig. 1(b), which include a self-homodyne interferometer using a 1-km optical fiber spool. The 1-km optical fiber spool, serving as a mechanical length reference, is based on the vibration-insensitive design described in Ref [30], but its longer length in this work results in a lower thermal noise limit. The spool, with dimensions of 100 mm in diameter and 20 mm in height, demonstrates a vibration sensitivity of approximately $10^{-10}$ [1/g]. The voltage signal detected from the BPD is fed back to the current modulator after passing through PI servos. The detailed feedback loop architecture is described in the Supplementary Material. The two separately stabilized CW lasers produce a beat note frequency of approximately 78.9 MHz. The frequency noise power spectral density was measured using a phase noise analyzer (53100A, Microsemi), and the frequency instability was calculated as the overlapping Allan deviation for averaging times greater than 0.001 seconds.

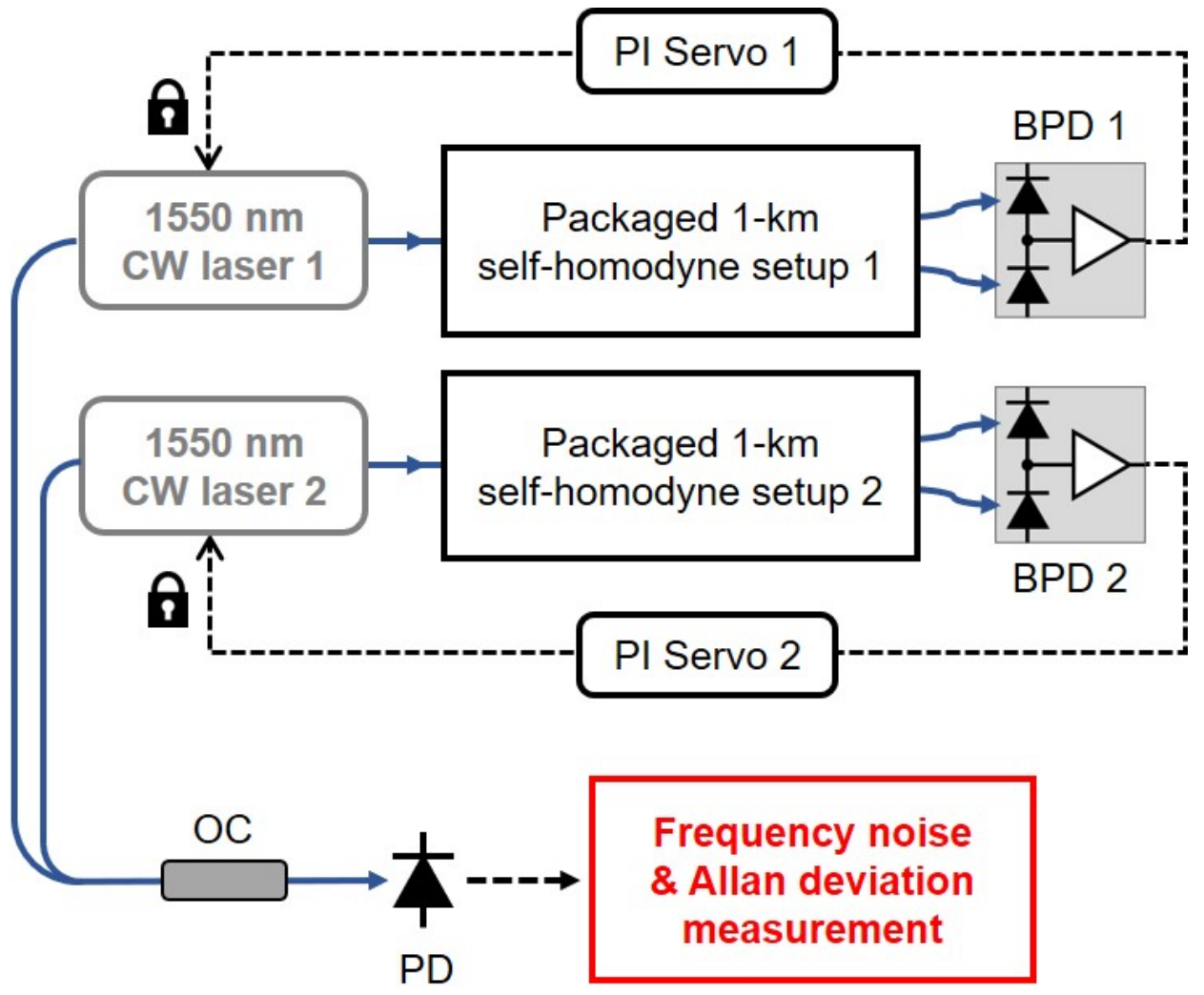

**Fig. 2** Schematic of the laser stabilization and noise measurement setup. Two 1550 nm fiber CW lasers are stabilized using separate all-fiber self-homodyne interferometers. The frequency noise and Allan deviation are then measured by analyzing the beat frequency between the two stabilized lasers.

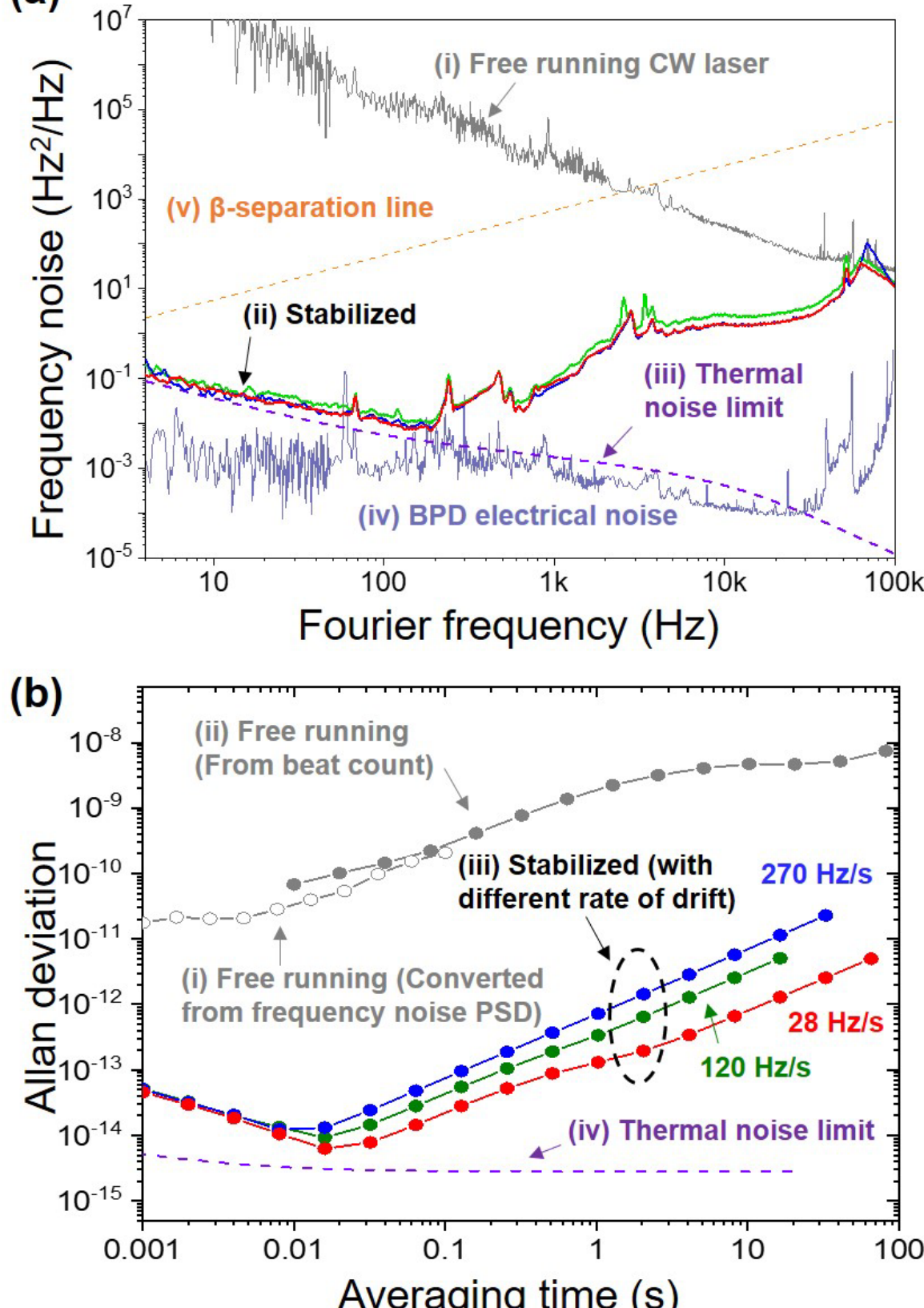

**Fig. 3** Frequency noise spectra and Allan deviation as results of CW lasers stabilization. (a) The frequency noise spectra. (i) free-running; (ii) stabilized; (iii) thermal-noise limit, which is the summantion of thermomechanical and thermo-refractive fiber noise [31]; and (iv) measured BPD electrical noise. (b) Optical frequency instability in terms of overlapping Allan deviation. (i) free-running frequency instability converted from frequency noise PSD (curve (i) in Fig. 4(a)); (ii) free-running frequency instability calculated from long-term beat measurement; (iii) overlapping Allan deviation of the stabilized CW lasers calculated from beat frequency time trace; and (iv) Thermal noise limit.

Figure 3(a) shows the frequency noise spectra for the CW lasers. When the laser is stabilized using the packaged system shown in Fig. 1(b), the frequency noise is reduced by up to $10^4$ times in the Fourier frequency range from 4 Hz to 10 kHz [curves (ii) in Fig. 3(a)] compared to the free-running CW laser [curve (i) in Fig. 3(a)]. The red, green, and blue curves represent different measurement trials, all of which demonstrate similar short-term frequency noise performance. Notably, during short-term noise measurements across all trials, the frequency noise level reaches the thermal noise limit [31,32] of the 1-km fiber round-trip (thus, 2-km effective length) in the 4-200 Hz Fourier frequency range, suggesting that an air-tight casing is sufficient to achieve the lowest frequency noise level without needing a vacuum system. The locking bandwidth in the Michelson interferometer with a 1-km fiber delay is limited to 100 kHz, which accounts for the prominent peak near 100 kHz in Fig. 3(a). Curve (iv) shows the BPD electrical noise, converted from the voltage noise measured from the in-loop BPD output after removing the 1-km spool from the long arm of the interferometer. The fact that this noise level is lower than the thermal noise of the round-trip of the 1-km optical fiber spool indicates that self-homodyne balanced detection is sufficient to stabilize the laser to the thermal-noise limit of a 2-km fiber. The β-separation line is well above the stabilized laser's frequency noise level, indicating that the linewidth within 0.1 seconds is less than 1 Hz [33]. (The integrated linewidth from a 1 Hz Fourier frequency is calculated as 12 Hz, which is relatively large due to longer-term thermal drift.) We also analyzed various types of residual noise, which are described in detail in the Supplementary Material.

Figure 3(b) illustrates the Allan deviation of the frequency instability [34]. The frequency instability reaches a minimum of $6.3\times10^{-15}$ at a 16-ms averaging time with minimal drift. The red, blue, and green curves represent the frequency instabilities corresponding to the trials shown in Fig. 3(a) with the same colors, demonstrating that although all trials produce similar short-term frequency noise PSD, the Allan deviation can differ based on the rate of fiber length drift. For each trial, the drift rates are 28 Hz/s, 120 Hz/s, and 270 Hz/s for the red, green, and blue curves, respectively, as shown in curve (iii) of Fig. 3(b).

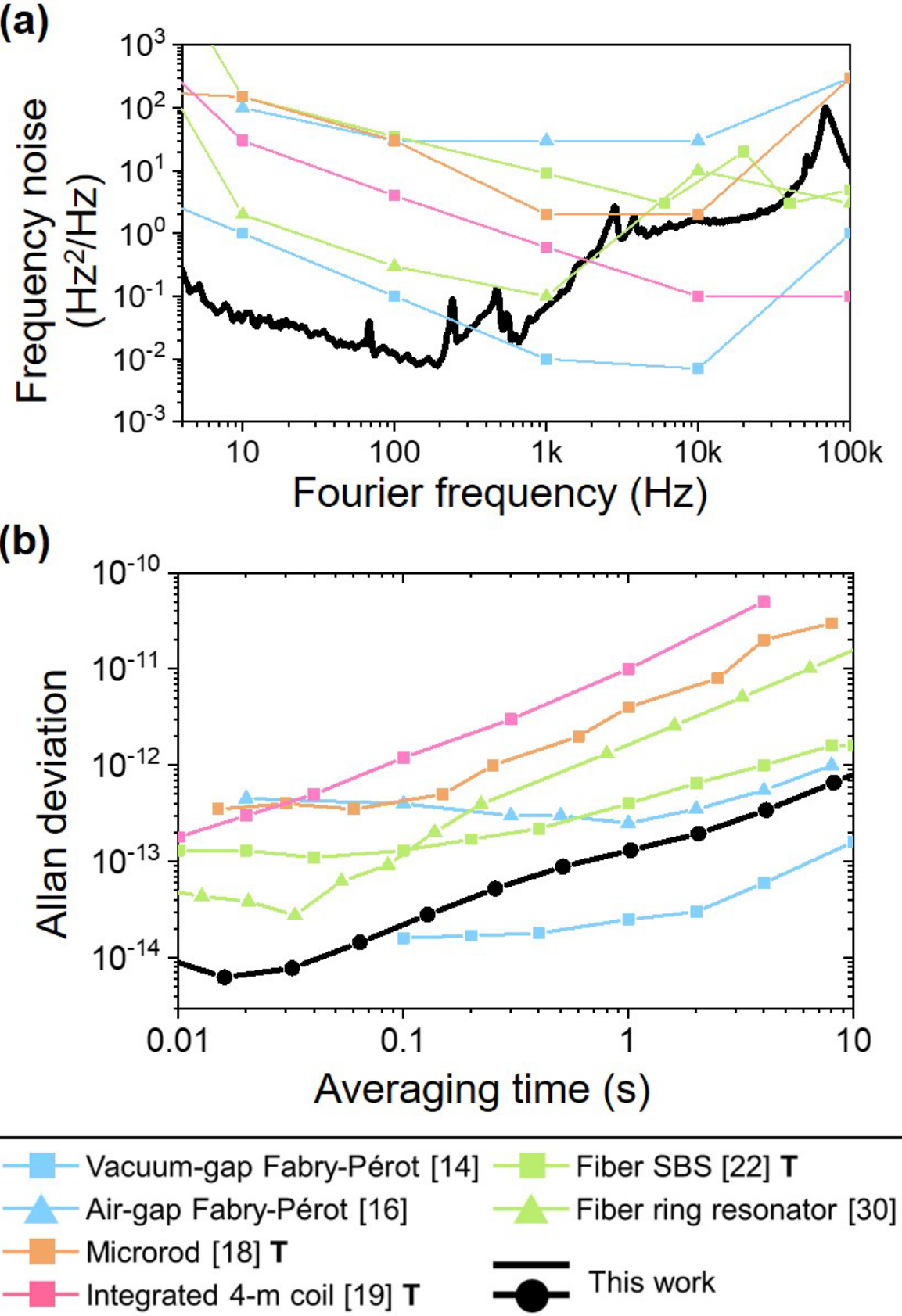

**Fig. 4** (a) Performance comparison of various non-vacuum and compact CW laser stabilization methods including our work. (a) Frequency noise spectrum. (b) Allan deviation. The capital T symbols in the legend indicate the setup is actively temperature-controlled.

Here we present a comparison of our work's performance with other recent non-vacuum compact CW laser stabilization platform results, as shown in Figure 4. Our work is compared with results from stabilization using vacuum-gap Fabry-Pérot cavities operated in air [14], air-gap compact Fabry-Pérot cavities [16], WGM resonators [18], integrated chip-scale resonators [19], and optical fiber delay lines [22,30]. As shown in Fig. 4, our frequency noise and Allan deviation are superior in the 4-100 Hz offset frequency or 10-100 ms averaging time range among other non-vacuum platforms. Achieving $10^{-15}$-level frequency instability with Fabry-Pérot cavities requires a vacuum and temperature control system, as noted in Refs [12,13]. These systems also require free-space coupling, which can be sensitive to external vibrations or impacts. Microrod and integrated chip resonators require sensitive and vulnerable tapered-fiber or butt coupling, making them difficult to use in harsh environments. Except for the fiber ring resonator configuration, all other platforms require the PDH method, which necessitates additional active optical components and RF components.

In summary, we have demonstrated a super-simple, compact, and alignment-free laser stabilization module based on an all-fiber self-homodyne interferometer. The optical frequency noise of a 1550-nm CW laser was stabilized to the thermal noise limit of the fiber delay. This resulted in minimum frequency instabilities of $6.3\times10^{-15}$ at 16-ms averaging time. The optical part of stabilization system is packaged in a compact aluminum case with dimension of $140\times180\times51$ mm$^3$, including vibration-insensitive 1-km fiber spool, which makes it highly suitable for field applications outside a laboratory environment.

**Funding.** National Research Foundation (NRF) of Korea (RS-2024-00334727 and RS-2021-NR060086), National Research Council of Science and Technology (NST) of Korea (CAP22061-000), Institute for Information and Communications Technology Promotion (IITP) of Korea (RS-2023-00223497)

**Disclosures.** I.J. and J.K.: KAIST (P).

**Data availability.** Data underlying the results presented in this paper are not publicly available at this time but may be obtained from the authors upon reasonable request.

## Full references